\documentclass[aps,twocolumn,showpacs,preprintnumbers,nofootinbib,prd,10pt,superscriptaddress,longbibliography,firstinits]{revtex4-2}

\makeatletter
\def\l@subsubsection#1#2{}
\def\l@subsubsubsection#1#2{}
\makeatother

\usepackage{graphicx,amssymb,amsmath,amsthm,amsfonts,epsfig,epsf,fixmath}
\usepackage{comment}
\usepackage{mathrsfs}
\usepackage[usenames]{color}
\usepackage{epstopdf}
\usepackage{dcolumn}
\usepackage{float}
\usepackage{latexsym}
\usepackage{rotating}
\usepackage[normalem]{ulem} 
\usepackage{longtable}

\usepackage{enumerate}
\usepackage{tensor,multirow}
\usepackage{url}
\usepackage{bm}
\usepackage[dvipsnames]{xcolor}
\usepackage[unicode]{hyperref}
\hypersetup{colorlinks=true, citecolor=MidnightBlue,
            linkcolor=Maroon, urlcolor=MidnightBlue, linktocpage=true}

\begin{document}

\title{
Semi-agnostic inspiral modeling: \\ the massless scalar case at leading post-Newtonian order}

\author{Dario Rossi}
\email{dario.rossi@phd.unipi.it}
\affiliation{Dipartimento di Fisica, Università di Pisa, Largo B. Pontecorvo 3, 56127 Pisa, Italy}
\affiliation{INFN, Sezione di Pisa, Largo B. Pontecorvo 3, 56127 Pisa, Italy}


\author{Leonardo Gualtieri}
\email{leonardo.gualtieri@unipi.it}
\affiliation{Dipartimento di Fisica, Università di Pisa, Largo B. Pontecorvo 3, 56127 Pisa, Italy}
\affiliation{INFN, Sezione di Pisa, Largo B. Pontecorvo 3, 56127 Pisa, Italy}

\author{Thomas~P.~Sotiriou}
\email{thomas.sotiriou@nottingham.ac.uk}
\address{Nottingham Centre of Gravity,
Nottingham NG7 2RD, United Kingdom}
\address{School of Mathematical Sciences, University of Nottingham,
University Park, Nottingham NG7 2RD, United Kingdom}
\address{School of Physics and Astronomy, University of Nottingham,
University Park, Nottingham NG7 2RD, United Kingdom}

\begin{abstract}

We introduce a {\em semi-agnostic} modeling framework for black hole binary inspirals beyond general relativity that fixes the field content without selecting a particular theory.
We expand in the post-Newtonian order and in powers of $l/L$, where $l$ is the characteristic length-scale that controls deviations from General Relativity and $L$ is the characteristic length-scale of the system. 
We focus on the case of a massless scalar field coupled with the metric, at leading post-Newtonian order.
Taking into account the scaling with $l/L$ of both the interaction terms and the quantities that characterise the sources, 
we find that the inspiral can be described in terms of
a small number of {\em universal} parameters. Such a parametrisation is suitable for obtaining combined bounds from multiple sources and performing searches for specific fields.
\end{abstract}

\maketitle

\section{Introduction}
Gravitational wave observations provide a new suite of tests of General Relativity (GR) in the strong field regime\,\cite{LIGOScientific:2025wao,LIGOScientific:2026oim}. One of these tests uses the inspiral part of the waveform, which is ideal for probing deviations from GR that endow compact objects with additional charges that source new polarizations. In such scenarios, the additional energy lost in these polarizations affects the orbital dynamics and hence leaves an imprint on the waveform, which accumulates over the inspiral. Indeed, inspiral tests currently give the tightest bounds on theories that predict that black holes (BHs) carry additional charges associated with massless fields \cite{Sanger:2024axs}.\footnote{See also \cite{Maselli:2021men,Speri:2024qak} for forecasted bounds from inspirals for very asymmetric binaries.}

The main approach for testing GR with GW inspirals is parametrised tests. This is done in a post-Newtonian (PN)  framework by adding a set of new parameters, $\delta \hat{\phi}_k$, which capture the deviation from GR at each PN order \cite{LIGOScientific:2026fcf}. Bounds on $\delta \hat{\phi}_k$ are obtained by only keeping one of the new parameters non-zero at a time. This is simple, efficient, and theory-agnostic, but at an expense. Consider a theoretical beyond-GR scenario (an alternative theory or a new field in some extension of the Standard Model) with a new coupling constant $\alpha$, for which GR is recovered as $\alpha\to 0$. Generically, $\alpha$ will control deviations in the structure of compact objects and in their dynamics. In such scenarios, several of the $\delta \hat{\phi}_k$ will typically be non-zero, not just one of them; they will depend on both $\alpha$ and the intrinsic source parameters (PN parameters encode source properties, both in GR and beyond,). The exact form of this dependence will be theory-specific. 

This example highlights the limitations of {\it ad hoc} parametrizations.  Bounds on  $\delta \hat{\phi}_k$ can be turned into bounds on $\alpha$ for given theory, but keeping only one of the $\delta \hat{\phi}_k$ at a time is not only inconsistent but introduces a truncation error. Moreover, the dependence of the $\delta \hat{\phi}_k$ on source parameters  means that one can only analyse single events. Combined bounds from multiple events, which would yield the most stringent tests of GR, can only be meaningful for {\em universal parameters}, such as $\alpha$. Finally, since the $\delta \hat{\phi}_k$ depend on both $\alpha$ and source parameters, the event that will give the tightest bound on $\alpha$ with theory-specific modeling will not generally be the one that gives the tightest bounds on the $\delta \hat{\phi}_k$ and vice versa (see e.g.\,\cite{Sanger:2024axs,Xie:2024xex}).
These limitations are well known and have motivated theory-specific tests as well, which tend to  outperform theory-agnostic ones. 
However, theory-specific tests can only be performed for a handful of well-motivated theories.  Parametrized tests clearly offer a more practical way to perform a first set of tests of GR for each event.

This is clearly not a binary choice as parametrised tests and theory-specific tests are highly complementary. But there is also a third way, which we aim to illustrate here, and we will call the {\em semi-agnostic framework}. The idea is that instead of selecting a particular theory, one can select a field content  and use principles of effective field theory (EFT) to create a modeling framework that captures all theories with that field content. 

At first, one is tempted to believe that such a framework will have an excessively large number of free parameters --- one for each new coupling constant associated with each interaction term --- to be of practical use. 
However, there are two main reasons why this initial expectation is not correct in general. The first relates to a common assumption
in EFTs: that these coupling constants, $\alpha_i$, will scale as $\tilde{\alpha}_i \, l^n$, where $l$ is a characteristic scale of the EFT, $\tilde{\alpha}_i$ are dimensionless ${\cal O}(1)$ parameters, and $n$ is determined by the dimensionality of the corresponding interaction term.
In geometric units $G=c=1$,  $l$ is a length and $n$ is positive. If $L$ is a characteristic length-scale of the system being studied and $l\ll L$, then $l/L$ becomes an additional natural perturbative parameter that ends up suppressing the effects of each interaction. Those interactions that come with a high enough $n$ become quantitatively irrelevant. Note that there can be technically natural EFTs with two different scales and interesting gravitational phenomenology, see e.g.~\cite{Noller:2019chl,Serra:2022pzl,Thaalba:2025lwe}. We will not consider this case here though, and stick to a single characteristic scale $l$.

The second simplification comes in a more subtle way: powers of $l/L$ also control the way deviations from GR are imprinted on stationary objects. The coefficients of a term in a PN expansion that acts as a waveform model are partially controlled by the parameters of the sources, and hence by the size of the deviations from GR for stationary objects. As a characteristic example, consider a theory with an additional massless scalar field that typically leads to dipolar emission. This comes at -1PN, with a magnitude controlled by the scalar charges of the objects composing the binary. If they carry zero charge, there is no dipolar emission. But the charge per unit mass,  which is dimensionless, has itself a scaling with $l/L$  --- indeed for a BH of mass $M$, it scales like $(l/M)^2$ --- and hence brings additional suppression in the $l/L$ expansion. 

It is the combination of the $l/L$ scaling of the interactions and the parameters characterising the GR deviations for the stationary objects that finally determines how important each interaction term is quantitatively. This rationale has been used in \cite{Maselli:2020zgv} to develop a modelling framework for extreme mass ratio inspiral (EMRI) and it has proved very efficient. 
This approach has allowed the development of waveform models for realistic orbits \cite{Barsanti:2022ana,Barsanti:2022vvl,DellaRocca:2024pnm,Barsanti:2026ulr,Gliorio:2026yvh},  produced the first LISA forecasts for EMRIs beyond GR \cite{Maselli:2021men,Speri:2024qak}, and it has been used to demonstrate that modelling to second order in the mass ratio can be done with only two new parameters \cite{Spiers:2023cva}. Ringdown modeling has also been shown to be very efficient in this semi-agnostic approach \cite{DAddario:2023erc,Lestingi:2025jyb,Asuncion:2026efk}. 

The goal of this paper is to introduce a similar approach into comparable mass binary inspiral modelling, where we will use as expansion parameters the dimensionless ratio $l/L$, and the post-Newtonian (PN) parameter $v/c$.
For concreteness, we fix the field content to be a metric and a massless scalar. Hence, we consider the Horndeski action~\cite{Horndeski:1974wa}, which is the most general action leading to second-order-in-time equations upon variation, and impose shift-symmetry, as it is the symmetry protecting the scalar from acquiring a mass from radiative corrections.
With these hypotheses, we will show how to implement the approach and how the simplifications discussed above arise already at leading PN order.  We will also discuss how this allows one to build an inspiral waveform model that uses universal parameters, without selecting a specific theory and hence allows one to meaningfully combine observations to obtain phenomenological bounds. 
 
Our analysis will act as a simple proof-of-principle and demonstration of  efficiency of the semi-agnostic approach in the context of comparable mass binaries. We note that our motivation and philosophy is similar to that used in Refs.~\cite{Payne:2024yhk,Bernard:2025dyh}, where a theory agnostic, EFT inspiral framework has also been introduced.
Our approach is different, but complementary. Indeed, while in\,\cite{Bernard:2025dyh} a wider choice of field content, and higher PN orders were considered, in the framework proposed here the 
dependence of the source parameters on the coupling constants of the theory is systematically exploited, leading to an expansion in $l/L$ that combines both this dependence and that of interaction terms in the field equations.

In Sec.\,\ref{sec:action} we introduce our general action and derive the corresponding field equations. In Sec.\,\ref{sec:scales} we introduce an EFT-inspired framework with scale separation, in which we solve the field equations; we obtain a description of the binary in terms of point particles characterized by their masses and scalar charges. In Sec.\,\ref{sec:2body_PN_Lagrangian} we derive the two-body PN Lagrangian, at leading PN order. Finally, in Sec.\,\ref{sec:concl} we draw our conclusions.
\section{Action and field equations}\label{sec:action}
As discussed in the Introduction, we will focus on the shift-symmetric subclass of Horndeski's theory~\cite{Sotiriou:2014pfa}, whose action is 
\begin{equation}
\label{eq:generic_shift-symmetric_HorndeskiAction}
    S=\frac{1}{2k}\sum_{i=2}^5 \int d^4x \sqrt{-g} \, \mathcal{L}_i + S_M\,,
\end{equation}
with
\begin{align}
    \mathcal{L}_2 =&  ~G_2(X)\,,\\
    \mathcal{L}_3 =&  -G_3(X)\Box \phi\,,\\
    \mathcal{L}_4 =&  ~G_4(X) R  + G_{4 X}[(\Box \phi)^2-(\nabla_{\mu} \nabla_\nu\phi)^2]\,,\\
    \mathcal{L}_5 =&  ~G_5(X) G_{\mu \nu} \nabla^{\mu} \nabla^\nu\phi  - \frac{G_{5 X}}{6}[(\Box \phi)^3\\
    &  \nonumber -3\Box\phi(\nabla_\mu \nabla_\nu \phi)^2+2(\nabla_{\mu} \nabla_\nu\phi)^3]\,,
\end{align}
where $k=8\pi G/c^4$, $X=-1/2\nabla_\mu\phi\nabla^\mu\phi$,  $(\nabla_\mu\nabla_\nu\phi)^2=\nabla_\mu\nabla_\nu\phi\nabla^\mu\nabla^\nu\phi$, $(\nabla_\mu\nabla_\nu\phi)^3=\nabla_\mu\nabla_\nu\phi\nabla^\nu\nabla^\gamma\phi\nabla_\gamma\nabla^\mu\phi$, $G_{i X}=\partial G_i/\partial X$, $R$ is the Ricci scalar, $G_{\mu\nu}$ the Einstein tensor, and $S_M$ is the matter action.

In order to choose a parametrization of the functions $G_i(X)$, we first note that the only coupling in the action\,\eqref{eq:generic_shift-symmetric_HorndeskiAction} which allows to evade the no-hair theorems of Ref.~\cite{Hui:2012qt}, and thus allows for stationary BHs different from those predicted by general relativity, is the Gauss-Bonnet coupling  $\alpha\phi\mathcal{G}$ \cite{Sotiriou:2013qea}, where $\mathcal{G}=R^2-4R^{\mu\nu}R_{\mu\nu}+R^{\mu\nu\alpha\beta}R_{\mu\nu\alpha\beta}$ is the Gauss-Bonnet invariant.
This coupling, together with the Einstein-Hilbert term and the canonical kinetic scalar term, leads to shift-symmetric scalar Gauss-Bonnet (sGB) gravity, which belongs to the Horndeski class. Indeed, it is obtained from the action\,\eqref{eq:generic_shift-symmetric_HorndeskiAction} with the choice $G_2=X$, $G_3=0$, $G_4=1$, $G_5=-4\alpha\ln|X|$\,\cite{Kobayashi:2011nu}. 

As discussed in\,\cite{Saravani:2019xwx}, for theories that respect locally Lorentz symmetry it is possible to write the Horndeski functions as those of sGB gravity plus analytic functions in $X\to0$, which can be Taylor-expanded in $X$: 
\begin{align}
        G_2(X)=&~X+\kappa X^2+\ldots \,,\nonumber\\
        G_3(X)=&-\sigma X+\ldots\,,\nonumber\\ 
        G_4(X)=&~1 + \gamma X+\ldots\,,\nonumber\\
        G_5(X)=&~-4\alpha\ln|X|+\eta X+\ldots\,.
\label{eq:Gi}
    \end{align}
    Keeping the leading order correction to sGB gravity for each of the $G_i$, 
leads to the action
\begin{align}
       S=&~\frac{1}{2k}\int d^4 x \sqrt{-g}\Bigg(R+X+\alpha \phi \mathcal{G} +\gamma G_{\mu\nu}\nabla^\mu\phi\nabla^\nu\phi \nonumber\\ 
       &  + \sigma X \Box \phi + \kappa X^2  + \eta\bigg(X G_{\mu\nu}\nabla^\mu\nabla^\nu\phi-\frac{1}{6}\big[(\Box\phi)^3\nonumber\\
&-3\Box\phi(\nabla_\mu\nabla_\nu\phi)^2+2(\nabla_\mu\nabla_\nu\phi)^3\big]\bigg) \Bigg)+S_M\,. 
 \label{eq:action}
\end{align}
This action is an extension of those considered in Ref.\,\cite{Thaalba:2022bnt}, where the term $\eta X$ was not included, and in Ref.\,\cite{DAddario:2023erc}, where the term $\kappa X^2$ was not included.\footnote{\label{foo_resc}Note that the action~\eqref{eq:action} is related to that used by Thaalba and collaborators in~\cite{Thaalba:2022bnt} by a minor change in notation: 
$\phi\to\sqrt{2}\phi$, $\alpha\to\sqrt{2}\alpha$, $\gamma\to\gamma$, $\sigma\to\sigma/\sqrt{2}$, $\kappa\to\kappa/2$.}
We will return to the contribution of the terms we have truncated in the following Sections.

With units such that $G=c=1$, the scalar field is dimensionless and the coupling constants $\alpha$, $\gamma$, $\sigma$, $\kappa$ have dimensions of length squared, while $\eta$ has dimensions of a length to the fourth power. 
From the action~\eqref{eq:action} we find the field equations for the metric:
\begin{align}
        G_{\mu\nu}=& ~\frac12\nabla_\mu\phi\nabla_\nu \phi-\frac{1}{4}g_{\mu\nu}(\nabla\phi)^2 + \alpha \, T^{(\alpha)}_{\mu\nu}+\gamma \, T^{(\gamma)}_{\mu\nu}\nonumber \\ & +\sigma \, T^{(\sigma)}_{\mu\nu}+\kappa \, T^{(\kappa)}_{\mu\nu}+\eta \, T^{(\eta)}_{\mu\nu}\,,
\label{eq:metric_field_equations}
\end{align}
where
\begin{align}
    T^{(\alpha)}_{\mu\nu} =&-\frac{1}{2g}g_{\mu(\rho}g_{\sigma)\nu} \epsilon^{\kappa\rho\alpha\beta}\epsilon^{\sigma\gamma\lambda\tau}R_{\lambda\tau\alpha\beta}\nabla_\gamma\nabla_\kappa\phi\,,\nonumber\\
    T^{(\gamma)}_{\mu\nu} =& -\frac12 R \nabla_\mu\phi\nabla_\nu\phi - 2G_{\gamma(\mu}\nabla_{\nu)}\phi\nabla^\gamma\phi -  R_{\mu\nu}X\nonumber\\ &-R_{\mu\alpha\nu\beta}\nabla^\alpha\phi\nabla^\beta\phi -(\nabla_\gamma\nabla_\mu\phi)(\nabla^\gamma\nabla_\nu\phi) \nonumber\\
    &   +\Box\phi(\nabla_\mu\nabla_\nu\phi) +\frac12g_{\mu\nu}\left[G_{\alpha\beta}\nabla^\alpha\phi\nabla^\beta\phi - (\Box\phi)^2\right.\nonumber\\
    &\left. +R_{\alpha\beta}\nabla^\alpha\phi\nabla^\beta\phi
    +(\nabla_\alpha\nabla_\beta\phi)^2 )\right]\, ,\nonumber\\
    T^{(\sigma)}_{\mu\nu} =&~\frac12 g_{\mu\nu}\nabla_\alpha\phi\nabla_\beta\phi \nabla^\alpha\nabla^\beta\phi + \frac12\nabla_\mu\phi\nabla_\nu\phi \Box\phi \nonumber\\
    &-  \nabla_\gamma \nabla_{(\mu}\phi\nabla_{\nu)}\phi\nabla^\gamma\phi\, , \nonumber\\
    T^{(\kappa)}_{\mu\nu} =& ~\frac12 g_{\mu\nu}\,X^2+\nabla_\mu\phi\nabla_\nu\phi \,X\,, \label{eq:defT}
\end{align}
and the explicit expression of $T^{(\eta)}_{\mu\nu}$ is shown in Appendix~\ref{appendix:field_equations}, and the equation for the scalar field,
\begin{align}
    \Box\phi=&-\alpha  \mathcal{G} + 2 \gamma G_{\mu\nu}\nabla^\mu\nabla^\nu\phi
    \nonumber\\
    &+\sigma\left[(\nabla_\mu\nabla_\nu\phi)^2 - (\Box\phi)^2+R_{\mu\nu}\nabla^\mu\phi\nabla^\nu\phi\right]\nonumber\\
    & + \,\kappa\left[2\nabla_\mu\nabla_\nu\phi\nabla^\mu\phi\nabla^\nu\phi-2 \Box\phi X\right] \,+\,\eta S_\eta(\phi)\,,
    \label{eq:scalar_field_equations}
\end{align}
where
\begin{align}
    S_\eta(\phi)=& - G_{\alpha\beta}\,\Box\phi\,\nabla^\beta\nabla^\alpha\phi+ G_{\alpha\beta}\,\nabla_\gamma\nabla^\alpha\phi\,\nabla^\gamma\nabla^\beta\phi \,\nonumber\\
    &+ G_{\gamma\delta}\,\nabla_\alpha\phi\,\nabla_\beta\phi\,R^{\beta\gamma\alpha\delta}   + R^{\alpha}{}_{\gamma}\,R^{\beta\gamma}\,\nabla_\alpha\phi\,\nabla_\beta\phi \nonumber\\
    &+ R^{\beta\gamma}\,\nabla_\beta\nabla^\alpha\phi\,\nabla_\gamma\nabla_\alpha\phi - R_{\beta}{}^{\gamma}\,\Box\phi\,\nabla_\gamma\nabla^\beta\phi\nonumber\\&
    - \frac12\nabla_\alpha\phi\,\nabla_\beta\phi\,R^{\alpha}{}_{\gamma\delta\varepsilon}\,R^{\beta\gamma\delta\varepsilon} + \nabla_\beta\nabla^\alpha\phi\,\nabla_\delta\nabla^\gamma\phi\,R_{\alpha\gamma}{}^{\beta\delta} .
\label{eq:defS}
\end{align}
Note that these are the field equations in vacuum, since we have not included the matter action $S_M$. 
\section{Separation of scales for an $N$-body system}\label{sec:scales}
The simplest and most effective approach to model a BH binary, computing the leading beyond-GR corrections to the emitted gravitational waveform, is the so-called {\it skeletonization approach}\,\cite{Eardley:1975fgi,Damour:1992we} in which a {\it separation of scales} is introduced; in modern terms, such approach can be considered as an EFT-inspired scheme. For the application of this approach to sGB gravity, see\,\cite{Yagi:2011xp,Julie:2019sab,Shiralilou:2021mfl,Almeida:2024cqz}.

In this approach we have a smaller scale, comparable to the dimensions of the bodies, and a much larger scale, comparable to their separation. In the smaller scale, each 
body can be described in its local frame by a stationary solution of the field equations in vacuum\,\eqref{eq:metric_field_equations},\,\eqref{eq:scalar_field_equations}. In the larger scale, 
each body can be treated as a point particle, moving in the spacetime obtained by solving the field equations in which we also include the 'matter' terms associated to the point particles themselves. The solution of these equations will depend on a set of parameters characterizing the point particles, and lead to a PN Lagrangian depending on the same parameters.
A crucial step in this construction is the {\it matching} between these two regimes, 
which yields the parameters of the point-particle system in terms of those characterizing the stationary BH solutions, i.e. their Arnowitt-Deser-Misner (ADM) masses and their scalar charges, which we will define below. 

In this section we shall follow these steps, finding the small-scale and large-scale
solutions  and performing the matching. 
\subsection{Stationary black hole solution}
We look for a spherically symmetric BH solution with line element, in Schwarzschild coordinates, given by
\begin{equation}
    ds^2=-A(r)dt^2+\frac{1}{B(r)}dr^2+r^2d\Omega^2.
\end{equation}
The solution can be found perturbatively, expanding the metric functions $A(r)$, $B(r)$ and the scalar field $\phi(r)$ in terms of the dimensionless coupling constants $\tilde{\alpha}=\alpha/r_h^2$,  $\tilde{\gamma}=\gamma/r_h^2$,  $\tilde{\sigma}=\sigma/r_h^2$,  $\tilde{\kappa}=\kappa/r_h^2$, $\tilde\eta=\eta/r_h^4$, where $r_h$ is the horizon radius of the BH, around the Schwarzschild metric with constant scalar field. We assume that all dimensionless constants are of the same scale,
\begin{equation}
\tilde\alpha\sim\tilde\gamma\sim\tilde\sigma\sim\tilde\kappa\sim\tilde\eta^{1/2}\,,
\end{equation}
which we assume to be small enough to perform the perturbative analysis. We thus introduce a bookkeeping parameter $q\ll1$, and redefine the dimensionless quantities as

\begin{equation}
\label{eq:coupling_constants}
\begin{aligned}
    \alpha=&~qr_h^2\tilde\alpha\,,\\
    \gamma=&~qr_h^2\tilde\gamma\,,\\
    \sigma=&~qr_h^2\tilde\sigma\,,\\
    \kappa=&~qr_h^2\tilde\kappa\,,\\
    \eta=&~q^2r_h^4\tilde\eta\,,
\end{aligned}
\end{equation}
such that $\tilde\alpha$, $\tilde\gamma$, $\tilde\sigma$, $\tilde\kappa$, $\tilde\eta$ are all of order $O(1)$. In other words, if $l$ is the length-scale of the couplings, $q\sim l^2/r^2_h$, i.e.~we have taken $L\sim r_h$. 

We then perform a perturbative expansion in $q$ around the Schwarzschild solution with constant scalar field,  as follows
\begin{subequations}
\label{eq:metricAndPhiExpansion}
\begin{align}
    A(r)=&\left(1-\frac{r_h}{r}\right)\left(1 + \sum_{n=1}^{\infty}A_n(r)q^n\right)^2\,,\\
    B(r)=&\left(1-\frac{r_h}{r}\right)\left(1 + \sum_{n=1}^{\infty}B_n(r)q^n\right)^{-2}\,,\\
    \phi(r)=&~\phi_0+\sum_{n=1}^{\infty}\phi_n(r)q^n\,.
\end{align}
\end{subequations}
Up to second order in $q$ we obtain
\begin{subequations}
\label{eq:second_order_microscopic_expansion}
\begin{align}
    A(r) =&~ \left(1-\frac{r_h}{r}\right)\left(1
         -q^2\frac{\tilde{\alpha }^2 }{15 r^6}\bigl(147 r^5 r_h+147 r^4 r_h^2\right. \nonumber \\
        & +137 r^3 r_h^3+7 r^2 r_h^4-50 r_h^6-26 r r_h^5\bigr)+\mathcal O(q ^3)\bigg)\,,\\
    B(r)=&~\left(1-\frac{r_h}{r}\right)\left(1-q^2\frac{\tilde{\alpha }^2}{15 r^6} \bigl(147 r^5 r_h+87 r^4r_h^2\right.\nonumber \\
    &+57 r^3 r_h^3-203 r^2 r_h^4-230 r_h^6-218 r r_h^5\bigr)+\mathcal O(q ^3)\bigg)\,,\\
    \phi(r)=&~\phi_0+q\frac{2 \tilde{\alpha } r_h \left(3 r r_h+2 r_h^2+6 r^2\right)}{3 r^3} + \mathcal O(q ^3)\,,
\end{align}
\end{subequations}
while the expressions up to $O(q^6)$ are given in Appendix~\ref{appendix:sixth_order_expressions}.

From Eqs.~\eqref{eq:second_order_microscopic_expansion} we see that at this order the only contributions arising from GR modifications are those associated to the GB coupling $\tilde\alpha$. As we will show below, the other corrections appear at higher order in $q$.

Transforming in harmonic coordinates $\{{\tilde x}^\mu\}$, such that $\partial_\mu(\sqrt{-\tilde g}{\tilde g}^{\mu\nu})=0$ (corresponding to $r=\tilde r+m+\mathcal{O}(1/\tilde r)$),
\begin{align}
\label{eq:asymptotic_microscopic_sol}
    \tilde g_{\mu\nu}=&~\eta_{\mu\nu}+\delta_{\mu\nu}\left(\frac{2m}{\tilde r}\right)+ \mathcal O\left(\frac{1}{\tilde r^2}\right)\,,\\
    \phi=&~\phi_0+ \frac{D}{\tilde r}+\mathcal O\left(\frac{1}{\tilde r^2}\right)\,,\label{eq:asymptotic_microscopic_sol2}    
\end{align}
where $m$ is the ADM mass and the coefficient $D$ is called \textit{scalar charge} of the body; note that $D$ is not a conserved N\"other charge, it is just a quantity characterizing the solution. By matching the asymptotic expansions of the metric near the horizon and at infinity and imposing regularity conditions order by order in $q$, one finds that the scalar charge is not a primary charge, i.e. that it is a function of the BH mass (and, for rotating BHs, of its angular momentum\,\cite{Kleihaus:2011tg,Kleihaus:2015aje,Delgado:2020rev}).
In the case of pure sGB gravity this is discussed e.g. in\,\cite{Mignemi:1992pm,Kanti:1995vq,Sotiriou:2014pfa,Julie:2019sab}, while in a forthcoming publication we consider 
a class of theories that does not satisfy shift symmetry
\cite{HerryDavid}.
In the case of shift-symmetric Horndeski gravity we find that
\begin{align}
    D=&~4 r_h \tilde{\alpha}q-\frac{4}{15} r_h \tilde{\alpha }^3 q^3-\frac{2 \tilde{\alpha }^4 r_h \left(829843 \tilde{\alpha }+41340 \tilde{\sigma }\right)}{17325}q^5 
    \nonumber\\
    &-\frac{16 \tilde{\alpha }^4 r_h \left(674536 \tilde{\alpha } \tilde{\gamma }-44204 \tilde{\alpha } \tilde{\kappa }-59895 \tilde{\eta }\right)}{45045}q^6\nonumber\\
    &+\mathcal O\left(q^7\right)\,.
\label{eq:D_6th}
\end{align}
We performed the expansion up to $O(q^6)$ since it is the lowest order in which contributions from all the terms included in the action\,\eqref{eq:action} appear in the expression of the scalar charge. In particular, we note that  the $\sigma$ term (i.e., the cubic Galileon, see e.g.\,\cite{Thaalba:2025lwe}) appears at order $q^5$, while the terms in $\gamma$, $\kappa$ and $\eta$ appear at order $q^6$.

The ADM mass is 
\begin{equation}
\begin{aligned}
    m=&~\frac{r_h}{2} \Bigg[1+\frac{49}{5}\tilde{\alpha}^2 q^2+\left(
\frac{36214}{1155}\tilde{\alpha}^3\tilde{\sigma}
+\frac{408253}{3850}\tilde{\alpha}^4
\right)q^4\\
&+\left(
\frac{7808224}{45045}\tilde{\alpha}^4\tilde{\gamma}
-\frac{31796}{3003}\tilde{\alpha}^3\tilde{\eta}
+\frac{2084944}{45045}\tilde{\alpha}^4\tilde{\kappa}
\right)q^5\\
&+\left(
-\frac{5083368}{85085}\tilde{\alpha}^4\tilde{\gamma}^2
+\frac{23677222}{85085}\tilde{\alpha}^4\tilde{\sigma}^2\right.\\
&\left.+\frac{169018655174}{98273175}\tilde{\alpha}^5\tilde{\sigma}+\frac{75242913669527}{26533757250}\tilde{\alpha}^6\right)q^6\\
&+\mathcal O\left(q^7\right)\Bigg]\,.
\end{aligned}
\label{eq:m_6th}
\end{equation}
Our results are in accordance with those of~\cite{Thaalba:2022bnt} (see footnote \ref{foo_resc} for a conversion among different notations). For pure sGB gravity Eq.\,\eqref{eq:D_6th} also agrees, modulo the different notations, with Eq.\,(C.4) of\,\cite{Julie:2019sab}, where the scalar charge is computed up to $\mathcal{O}(q^4)$.


\subsection{Skeletonization and matching}
Following~\cite{Eardley:1975fgi,Damour:1992we} we replace the matter action $S_M$ with the skeletonized action of $N$ bodies, of the form (see also\,\cite{Julie:2019sab,Shiralilou:2021mfl})
\begin{equation}
\label{eq:PP_action}
    S_m=-\sum_A\int m_A(\phi)\sqrt{-g_{\mu\nu}\frac{dx^\mu_A}{d\tau}\frac{dx^\nu_A}{d\tau}}d\tau\,,
\end{equation}
where $A=1,\dots,N$, $x_A^\mu(\tau)$ is the worldline of the $A$-th particle, and $m_A(\phi)$ is a function characterizing the coupling of the particle with the scalar field, which will be determined through the matching with the microscopic solution.

 The corresponding field equations are:
\begin{align}
    G_{\mu\nu}=& ~\frac12\nabla_\mu\phi\nabla_\nu \phi-\frac{1}{4}g_{\mu\nu}(\nabla\phi)^2 + \alpha \, T^{(\alpha)}_{\mu\nu}+\gamma \, T^{(\gamma)}_{\mu\nu}\nonumber \\ 
    & +\sigma \, T^{(\sigma)}_{\mu\nu}+\kappa \, T^{(\kappa)}_{\mu\nu}+\eta\,T^{(\eta)}_{\mu\nu}\nonumber\\
    & +8\pi\sum_A\int d\tau\, m_A(\phi)\frac{\delta^{(4)}(x-x_A(\tau))}{\sqrt{-g}}\frac{d x^A_\mu}{d\tau}\frac{d x^A_\nu}{d\tau}\,,\nonumber\\
    &\label{eq:macroscopic_metric}\\
    \Box\phi=&-\alpha \mathcal{G} + 2 \gamma G_{\mu\nu}\nabla^\mu\nabla^\nu\phi \nonumber\\ 
    &  +\sigma\big[(\nabla_\mu\nabla_\nu\phi)^2 - (\Box\phi)^2+R_{\mu\nu}\nabla^\mu\phi\nabla^\nu\phi\big]\nonumber \nonumber\\
    & +\kappa\left[2\nabla_\mu\nabla_\nu\phi\nabla^\mu\phi\nabla^\nu\phi-2\Box\phi X\right]  \nonumber\\
    &   + \,\eta S_\eta(\phi) + 16\pi\sum_A\int d\tau\,m_A'(\phi)\frac{\delta^{(4)}(x-x_A(\tau))}{\sqrt{-g}}\,,\nonumber\\
   \label{eq:macroscopic_scalar}
\end{align}
where the prime denotes differentiation with respect to the argument. We solve these equations perturbatively around Minkowski spacetime and constant scalar field. In the rest frame of particle $A$, using harmonic coordinates so that $\partial_\mu(\sqrt{-\tilde g}\tilde g^{\mu\nu})=0$, we obtain
\begin{align}
    \tilde g_{\mu\nu}=&~\eta_{\mu\nu}+\delta_{\mu\nu}\left(\frac{2m_A(\phi_0)}{\tilde r}\right)+\mathcal{O}\left(\frac{1}{\tilde r^2}\right)\,,\nonumber\\
    \phi=&~\phi_0-\frac{4}{\tilde r}m_A'(\phi_0)+\mathcal{O}\left(\frac{1}{\tilde r^2}\right)\,.
    \label{eq:macroscopic_expansion}
\end{align}
Note that the $1/{\tilde r}$ terms in the expansion\,\eqref{eq:macroscopic_expansion} do not depend on the GR modification, i.e. on $\alpha$, $\sigma$, $\gamma$, $\kappa$, $\eta$.
This happens for two different reasons: the contributions coming from the $G_i$ terms of Eq.~\eqref{eq:Gi} (i) vanish at first order in the perturbation around the Minkowski, constant scalar field background, and (ii) vanish at infinity at higher order in $1/{\tilde r}$ (indeed, Eq.\,\eqref{eq:macroscopic_expansion} implies that $\nabla_\mu\phi\sim 1/{\tilde r}^2$).

By comparing Eq.~\eqref{eq:macroscopic_expansion} with Eqs.~\eqref{eq:asymptotic_microscopic_sol}-\eqref{eq:asymptotic_microscopic_sol2}, we obtain the {\it matching conditions}
\begin{align}
    &m_A(\phi_0)=m\,,\label{eq:match_mass}\\
    &m_A'(\phi_0)=-D/4\,,\label{eq:match_charge}
\end{align}
relating the function $m_A$ and its derivative at $\phi_0$ with the ADM mass of the BH and its scalar charge. 
We remind that the expression of the scalar charge in terms of the mass and of the couplings of the theory is given in Eqs.\,\eqref{eq:D_6th} and \eqref{eq:m_6th}. 

We define $m_A^0=m_A(\phi_0)$, and $\alpha_A^0=m_A'(\phi_0)/m_A^0$, which we call {\it BH sensitivity}~\cite{Eardley:1975fgi,Damour:1992we}. This quantity, which characterizes the dipolar emission of the system, is:
\begin{align}
\alpha_A^0=&~ -2 q \tilde{\alpha }+\frac{296}{15} q^3 \tilde{\alpha }^3+\frac{16 q^5 \left(70485 \tilde{\alpha }^4 \tilde{\sigma }+72106 \tilde{\alpha }^5\right)}{17325}\nonumber\\
&+\frac{16 q^6 \left(1313296 \tilde{\alpha }^5 \tilde{\gamma }-89565 \tilde{\alpha }^4 \tilde{\eta }+238516 \tilde{\alpha }^5 \tilde{\kappa }\right)}{45045}\nonumber\\
&+\mathcal O\left(q^7\right)\,.    \label{eq:sensitivity}
\end{align}
We note that the interactions in the Lagrangian different from the GB term do not affect the sensitivity up to $\mathcal{O}({q}^5)$. 

The expression for the sensitivity, Eq.\,\eqref{eq:sensitivity}, does not include 
the contributions from the terms in eq.~\eqref{eq:Gi} that we truncated. However,  for each of the $G_i$ the terms we neglected are suppressed by at least a factor of $q$ with respect to the leading order correction that we have kept. This implies that eq.~\eqref{eq:sensitivity} is robust to order $q^5$ at least, which is sufficient for our purposes.

\section{Two-body PN Lagrangian}
\label{sec:2body_PN_Lagrangian}
Let us consider the Lagrangian describing the motion of the $N$ bodies at leading PN order\,\cite{Damour:1992we,Julie:2019sab,Shiralilou:2021mfl}, i.e. at leading order in the expansion in
$m/r\sim v^2\ll1$, with $r$ representing the distance between the two bodies, and $v$ their relative velocity. We expand around Minkowski spacetime, $\eta_{\mu\nu}$, and a constant scalar field, $\phi_0$, introducing global
harmonic coordinates ($\partial_\mu(\sqrt{-g}g^{\mu\nu})=0$), such that $\sqrt{-g}g^{\mu\nu}=\eta^{\mu\nu}+h^{\mu\nu}$ and $\phi=\phi_0+\delta\phi$. The separation $r$ is defined in this frame. The leading relativistic corrections to the metric, $h^{\mu\nu}$, and to the scalar field, $\delta\phi$, are of order
\begin{subequations}
    \begin{align}
        &h^{00}\sim\mathcal{O}(v^2)\,,\\
        &h^{0i}\sim\mathcal{O}(v^3)\,,\\        
        &h^{ij}\sim\mathcal{O}(v^4)\,,\\
        &\delta \phi\sim\mathcal{O}(v^2)\,.
    \end{align}
\end{subequations}
The function $m_A(\phi)$, in this expansion, is then
\begin{equation}
\begin{aligned}
    \label{eq:mass_function}
    m_A(\phi)=&~m_A^0 \bigg[1+\alpha_A^0\delta\phi
+\mathcal{O}(\delta\phi^2)    \bigg]\,.
    \end{aligned}
\end{equation}
In the following we consider a system of two bodies, denoted by $A=1$ and $B=2$
. In the PN expansion of field equations, the terms $v^n$ are of PN order $(n-2)/2$; thus, the terms $\mathcal O(v^2)$ are of leading PN order, i.e. 0PN. Each derivative of the scalar field or of the metric leads to a term of order $v$ or higher.\footnote{Time derivatives lead to terms $\mathcal O(v)$ while space derivatives lead to terms $\mathcal O(m/r)=\mathcal O(v^2)$. Note that the terms with higher derivatives in the field equations have coefficients $\alpha\sim m^2$, $\sigma\sim m^2$, etc.} We can see that
all contributions coming from the scalar field sector in the field equations\,\eqref{eq:macroscopic_metric}, \eqref{eq:macroscopic_scalar}, with the definitions\,\eqref{eq:defT}, \eqref{eq:defS}, \eqref{eq:defTeta}, are of order $v^4$ or higher, therefore they are 1PN or higher-PN order corrections. 
Therefore, for the theory we considered, we can write the two-body Lagrangian at 0PN order as in scalar-tensor gravity without Horndeski couplings, i.e.\,\cite{Damour:1992we} 
\begin{align}
    L_{AB}=&-m_A^0-m_B^0+\frac12m_A^0v_A^2+\frac12m_B^0v_B^2\nonumber\\
    &+\frac{m_A^0m_B^0}{r}(1+4\alpha_A^0\alpha_B^0)+\mathcal{O}(v^4)\,.\label{eq:LAB}
    \end{align}
This is the same expression derived e.g. in~\cite{Julie:2019sab} for sGB gravity. 

Remarkably, the GB terms and, more generally, the Horndeski terms considered in this paper do affect the Lagrangian\,\eqref{eq:LAB} only through the expression of the sensitivity, Eq.\,\eqref{eq:sensitivity}.\footnote{This statement continues to be true for interesting models beyond shift-symmetry, and can allow one to translate bounds between theories \cite{HerryDavid}.} They directly enter in the Lagrangian at higher PN orders. 
We remind that due to the matching conditions\,\eqref{eq:match_mass}, \eqref{eq:match_charge}, the quantities $m_{A,B}^0$ and $\alpha_{A,B}^0$ are the ADM masses and sensitivities, respectively, of the two BHs, and
depend on the coefficients $\tilde\alpha$, $\tilde\sigma$, $\tilde\gamma$, $\tilde\kappa$, $\tilde\eta$ through Eq.\,\eqref{eq:sensitivity}. This conclusion can be easily extended to higher-order terms in the $X$-expansion\,\eqref{eq:Gi}, since each additional $X$ term corresponds to an additional 3PN order.

It is  worth clarifying that the parameter $q$ in eq.~\eqref{eq:sensitivity} 
is defined 
in terms of one of the available characteristic length-scales, e.g. the horizon radius $r_h$ of one of the two BHs or, say, the total mass. However, for comparable mass binaries all of these length-scales are comparable and the choice simply determines numerical coefficients. Choosing the largest of the length-scales guarantees that these coefficients are smaller than one.

Finally, we remark  that we are assuming that the background scalar field $\phi_0$ is constant during the inspiral, and thus the scalar charges and the ADM masses are constant as well. As discussed in\,\cite{Julie:2019sab,Julie:2022huo}, this is generally  not true. However, the evolution of these quantities is a higher-PN phenomenon, which can consistently be neglected in the $0$PN analysis of this paper.
\section{Discussion}\label{sec:concl}

This paper is the first step in the development of a semi-agnostic 
framework to model BH binary inspirals, in which the field content is fixed while the theory is left unspecified. 
As a first concrete example we have considered the case of a shift-symmetric, massless scalar field. 
With the further assumption that all the dimensionful coupling constants depend on a single fundamental length $l$, we have proceeded to model the binary using a double expansion in post-Newtonian orders and in 
$q\sim l^2/L^2$, 
where $L$ is a characteristic length-scale of the binary.  

To model the BH binary, we have performed an EFT-inspired separation of scales. Each body is described, in the smaller scale, as a stationary BH, and in the larger scale as a point particle. By matching the two descriptions, the point particles are characterized by their masses and scalar charges, and thus by their sensitivities. The coupling between the scalar and the Gauss-Bonnet invariant contributes to the sensitivity at $O(q)$, while corrections from other couplings come at $O(q^5)$.
We have then derived a two-body PN Lagrangian at leading PN order, finding that it has the standard form of scalar-tensor gravity\,\cite{Damour:1992we} and the various shift-symmetric interaction couplings only appear through the BH sensitivities. Therefore, the Gauss-Bonnet coupling gives the leading contribution to the 0PN inspiral, while any other contribution is suppressed by a factor of $q^5$ at least.

Our result explicitly demonstrates that at 0PN, the combination of the sensitivities acts as a single new parameter,
while Eq.~\eqref{eq:sensitivity}  can be used to express this parameter in terms of $q$ and the dimensionless couplings. 
At face value the
parametrisation in terms of the sensitivities is more compact, while expressing it in terms of coupling constants introduces 
a larger number of
free parameters. Up to here this seems to confirm the usual trade-off between a theory-agnostic parametrization whose parameters have an unknown dependence on source parameters --- and hence cannot meaningfully be used for combined bounds from different observations --- and a theory-specific parametrization that used universal couplings at the expense of having more free parameters. But the scaling with $q$ is key in changing this perspective. 
LVK bounds on sGB gravity imply that $q\lesssim 0.1$\,\cite{Sanger:2024axs}. The fact that the other couplings in Eq.~\eqref{eq:sensitivity} are suppressed by at least $q^5$ implies that this bound is robust at 0PN level. Hence these couplings can safely be ignored quantitatively for a 0PN waveform. Under this approximation, one can think of $l$ as a single {\em universal} parameter,\footnote{The dimensionless Gauss-Bonnet coupling $\tilde{\alpha}$ (which is ${\cal O}(1)$) can be absorbed in a redefinition of $l$ without loss of generality.} and have an accurate enough waveform that can be used to obtain combined bounds from multiple observations. 

We expect this to be a powerful tool when pushing this framework to higher orders. The two-body PN Lagrangian will receive contributions from new couplings in two ways: (i) an interaction term can contribute directly a new higher PN term in eq.~\eqref{eq:LAB}; (ii) the coefficient of a higher PN term can depend on a new coupling via the effect the latter has on the multipoles of the sources. In both cases, these contributions will be suppressed by some power of $q$ and the exponent can be computed within the
framework we have laid out here. Crucially, at any given PN order, only low enough powers of $q$ will be quantitatively relevant and the error in dropping them is quantifiable. At the same time, the dependence of $q$ and of the dimensionless couplings on the source parameters is known. The parametrization is then expected to depend on a small number of quantitative relevant, universal free parameters.

This demonstrates a point of principle and acts as a first step towards developing a semi-agnostic framework 
for new field searches with BH inspirals. The next steps will be the extension to higher PN orders, and to rotating BHs. These will be the subjects of  forthcoming publications.
\acknowledgements
We thank Emanuele Berti, Herry Lalus, David Trestini and Enrico Trincherini for useful discussions. This project has received funding from the European Union’s Horizon Europe research and innovation programme under the Marie Skłodowska-Curie Staff Exchanges grant agreement No. 101299389 (STRONG).
T.P.S. acknowledges partial support from the STFC Consolidated Grant nos. ST/V005596/1,  ST/X000672/1, and UKRI2492. 
\\

\appendix

\section{Explicit expression of $T^{(\eta)}_{\mu\nu}$}
\label{appendix:field_equations}
Here we write the explicit expression of the term $T^{(\eta)}_{\mu\nu}$ appearing in Eq.\,\eqref{eq:metric_field_equations}:
\begin{equation}
\begin{aligned}
T^{(\eta)}_{\mu\nu}=&~\frac{1}{2}\Big[ 
- 2R^{\gamma}{}_{(\mu}\nabla_{\nu)}\nabla^\delta\phi \nabla_\delta\phi\nabla_\gamma\phi + G_{\mu\nu} \nabla_\gamma\phi \nabla^\delta\nabla^\gamma\phi \nabla_\delta\phi 
\\
& - 2R^{\delta\varepsilon\gamma}{}_{(\mu} \nabla_{\nu)}\nabla_\varepsilon\phi \nabla_\gamma\phi \nabla_\delta\phi
-2 \nabla_\gamma\nabla_\mu\phi \nabla_\delta\nabla_\nu\phi \nabla^\delta\nabla^\gamma\phi 
\\
&+ 2 R_{\mu}{}^{(\delta\varepsilon)}{}_{\nu} \nabla_\gamma\phi \nabla_\varepsilon\nabla^\gamma\phi \nabla_\delta\phi
- g_{\mu\nu} R^{\gamma\delta} \nabla_\gamma\phi \nabla_\delta\phi \Box\phi\\
&+ R_{\mu}{}^{\gamma}{}_{\nu}{}^{\delta} \nabla_\gamma\phi \nabla_\delta\phi\Box\phi
- 2g_{\mu\nu} R^{\gamma}{}_{\varepsilon} \nabla_\gamma\phi \nabla^\varepsilon X \\
&+ g_{\mu\nu} R^{\gamma}{}_{\varepsilon}{}^{\delta\zeta} \nabla_\gamma\phi \nabla_\zeta\nabla^\varepsilon\phi \nabla_\delta\phi
+2 R_{(\mu}{}^{\gamma} \nabla_{\nu)}\phi \nabla_\gamma\phi \Box\phi\\
&
+2 \nabla_\gamma\nabla_\mu\phi \nabla^\gamma\nabla_\nu\phi \Box\phi - 2 R^{\gamma}{}_{\delta} \nabla_\gamma\phi \nabla^\delta\nabla_{(\mu}\phi \nabla_{\nu)}\phi \\
& +\nabla_\nu\nabla_\mu\phi \Big(R^{\gamma\delta}\nabla_\gamma\phi \nabla_\delta\phi
-(\Box\phi)^2
+(\nabla_\delta\nabla_\gamma\phi)^2\Big)\\
&+2 R^{\gamma\varepsilon}{}_{\delta(\mu}\nabla_{\nu)}\phi \nabla_\gamma\phi \nabla_\varepsilon\nabla^\delta\phi
 +\Big( G_{\gamma\delta} \nabla_\mu\phi \nabla_\nu\phi 
\\
&- G_{\nu\delta} \nabla_\mu\phi \nabla_\gamma\phi - G_{\mu\delta} \nabla_\nu\phi \nabla_\gamma\phi\Big)\nabla^\delta\nabla^\gamma\phi\\
&+\frac13 g_{\mu\nu} \left((\Box\phi)^3+2 (\nabla^\delta\nabla^\gamma\phi)^3 - 3 \Box\phi (\nabla_\varepsilon\nabla_\delta\phi)^2\right)
\Big]\,.\label{eq:defTeta}
\end{aligned}
\end{equation}
\section{Expansion of the metric and scalar field in terms of the coupling constants}
\label{appendix:sixth_order_expressions}
Here we provide the expressions of the functions $A_n(r)$, $B_n(r)$ and $\phi_n(r)$ in Eqs.\,\eqref{eq:metricAndPhiExpansion} up to sixth order in $q$:
\begin{widetext}
\begin{align*}
A_1(r)=&~0\,,\\[2ex]
A_2(r)=&~-\frac{\tilde{\alpha }^2 r_h}{30 r^6} \left(147 r^4 r_h+137 r^3 r_h^2+7 r^2 r_h^3-26 r r_h^4-50 r_h^5+147 r^5\right)\,,\\[2ex]
A_3(r)=&~0\,,\\[2ex]
A_4(r)=&~-\frac{\tilde{\alpha }^3 r_h}{138600 r^{12}} \left(\tilde{\alpha } \left(9012447 r^{10} r_h+10229740 r^9 r_h^2
+5371733 r^8 r_h^3+1453742 r^7 r_h^4-3023423 r^6 r_h^5\right.\right.\\
&\left.\left.+9529964 r^5 r_h^6+11885251 r^4 r_h^7+15605968 r^3 r_h^8
-1270584 r^2 r_h^9-4684400 r r_h^{10}-8354500 r_h^{11}\right.\right.\\
&\left.\left.+7348554 r^{11}\right)+120 \tilde{\sigma } \left(18107 r^{10} r_h+18107 r^9 r_h^2
+18107 r^8 r_h^3+18107 r^7 r_h^4+18107 r^6 r_h^5-3013 r^5 r_h^6\right.\right.\nonumber\\
&\left.\left.-11263 r^4 r_h^7-20118 r^3 r_h^8+210 r^2 r_h^9+3150 r r_h^{10}
+6930 r_h^{11}+18107 r^{11}\right)\right)\,.\\[2ex]
A_5(r)=&~-\frac{2 \tilde{\alpha }^3 r_h}{45045 r^{15}} \left(4 \tilde{\alpha } \left(\tilde{\gamma } \left(488014 r^{13} r_h+488014 r^{12} r_h^2+488014 r^{11} r_h^3+488014 r^{10} r_h^4+494020 r^9 r_h^5\right.\right.\right.\nonumber\\
&\left.\left.\left.+395350 r^8 r_h^6+331000 r^7 r_h^7+250920 r^6 r_h^8+809478 r^5 r_h^9+742320 r^4 r_h^{10}+657690 r^3 r_h^{11}-270930 r^2 r_h^{12}\right.\right.\right.\nonumber\\
&\left.\left.\left.-317460 r r_h^{13}-330330 r_h^{14}+488014 r^{14}\right)+\tilde{\kappa } \left(130309 r^{13} r_h+130309 r^{12} r_h^2+130309 r^{11} r_h^3\right.\right.\right.\nonumber\\
&\left.\left.\left.+130309 r^{10} r_h^4+136315 r^9 r_h^5+140605 r^8 r_h^6-39575 r^7 r_h^7-254790 r^6 r_h^8-489024 r^5 r_h^9-339966 r^4 r_h^{10}\right.\right.\right.\nonumber\\
&\left.\left.\left.-129756 r^3 r_h^{11}+126654 r^2 r_h^{12}+109824 r r_h^{13}+66066 r_h^{14}+130309 r^{14}\right)\right)-3 \tilde{\eta } \left(39745 r^{13} r_h\right.\right.\nonumber\\
&\left.\left.+39745 r^{12} r_h^2+39745 r^{11} r_h^3+39745 r^{10} r_h^4+39745 r^9 r_h^5+28305 r^8 r_h^6+41175 r^7 r_h^7+56190 r^6 r_h^8\right.\right.\nonumber\\
&\left.\left.+144278 r^5 r_h^9+120254 r^4 r_h^{10}+87494 r^3 r_h^{11}-44176 r^2 r_h^{12}-47476 r r_h^{13}-44044 r_h^{14}+39745 r^{14}\right)\right)\,,\\[2ex]
A_6(r)=&~-\frac{\tilde{\alpha }^4 r_h}{212270058000 r^{18}} \left(360 \tilde{\alpha } \tilde{\sigma } \left(552350607915 r^{16} r_h+591951702448 r^{15} r_h^2+551172486971 r^{14} r_h^3\right.\right.\nonumber\\
&\left.\left.+508095227746 r^{13} r_h^4+451534021044 r^{12} r_h^5+409096467314 r^{11} r_h^6+280442404634 r^{10} r_h^7\right.\right.\nonumber\\
&\left.\left.+75334121254 r^9 r_h^8+614601983387 r^8 r_h^9+793978847144 r^7 r_h^{10}+1168330107791 r^6 r_h^{11}-241930790332 r^5 r_h^{12}\right.\right.\nonumber\\
&\left.\left.-667217470920 r^4 r_h^{13}-1274066357564 r^3 r_h^{14}-21991539570 r^2 r_h^{15}+165464699400 r r_h^{16}+425719394100 r_h^{17}\right.\right.\nonumber\\
&\left.\left.+507055965522 r^{17}\right)+\tilde{\alpha }^2 \left(356118856125446 r^{16} r_h+419731003122845 r^{15} r_h^2+405374130512765 r^{14} r_h^3\right.\right.\nonumber\\
&\left.\left.+309917067582536 r^{13} r_h^4+147317262693014 r^{12} r_h^5+344230900527356 r^{11} r_h^6+404410107327926 r^{10} r_h^7\right.\right.\nonumber\\
&\left.\left.+592099761327774 r^9 r_h^8-246002412581226 r^8 r_h^9-428040919721625 r^7 r_h^{10}-827206519007715 r^6 r_h^{11}\right.\right.\nonumber\\
&\left.\left.+936949008258186 r^5 r_h^{12}+1191470253711738 r^4 r_h^{13}+1641580998693636 r^3 r_h^{14}-198211568454900 r^2 r_h^{15}\right.\right.\nonumber\\
&\left.\left.-477310919085000 r r_h^{16}-804876957885000 r_h^{17}+300971654678108 r^{17}\right)-831600 \left(4 r \tilde{\gamma }^2 \left(1906263 r^{15} r_h\right.\right.\right.\nonumber\\
&\left.\left.\left.+1906263 r^{14} r_h^2+1906263 r^{13} r_h^3+1906263 r^{12} r_h^4+1906263 r^{11} r_h^5+1906263 r^{10} r_h^6+2052123 r^9 r_h^7\right.\right.\right.\nonumber\\
&\left.\left.\left.+2222293 r^8 r_h^8-908835 r^7 r_h^9-4844403 r^6 r_h^{10}-9299763 r^5 r_h^{11}-6550863 r^4 r_h^{12}-2534103 r^3 r_h^{13}\right.\right.\right.\nonumber\\
&\left.\left.\left.+2483481 r^2 r_h^{14}+2177175 r r_h^{15}+1321320 r_h^{16}+1906263 r^{16}\right)-\tilde{\sigma }^2 \left(35515833 r^{16} r_h+35515833 r^{15} r_h^2\right.\right.\right.\nonumber\\
&\left.\left.\left.+35515833 r^{14} r_h^3+35515833 r^{13} r_h^4+35515833 r^{12} r_h^5+35515833 r^{11} r_h^6+33765513 r^{10} r_h^7+33084833 r^9 r_h^8\right.\right.\right.\nonumber\\
&\left.\left.\left.-18919119 r^8 r_h^9-46263891 r^7 r_h^{10}-81349851 r^6 r_h^{11}-3359631 r^5 r_h^{12}+20875569 r^4 r_h^{13}+55852797 r^3 r_h^{14}\right.\right.\right.\nonumber\\
&\left.\left.\left.+3678675 r^2 r_h^{15}-2942940 r r_h^{16}-13273260 r_h^{17}+35515833 r^{17}\right)\right)\right)\,;
\end{align*}
\begin{align*}
B_1(r)=&~0\,,\\[2ex]
B_2(r)=&~\frac{\tilde{\alpha }^2 r_h}{30 r^6} \left(87 r^4 r_h+57 r^3 r_h^2-203 r^2 r_h^3-218 r r_h^4-230 r_h^5+147 r^5\right)\,,\\[2ex]
B_3(r)=&~\frac{8 \tilde{\alpha }^2 \tilde{\gamma }} {r^9}\left(r-r_h\right) r_h^4 \left(r r_h+r_h^2+r^2\right){}^2\,,\\[2ex]
B_4(r)=&~\frac{\tilde{\alpha }^3 r_h}{138600 r^{12} \left(r-r_h\right)} \left(r_h \left(-422961 r^{10} \tilde{\alpha } r_h-12880021 r^9 \tilde{\alpha } r_h^2-14661801 r^8 \tilde{\alpha } r_h^3\right.\right.\nonumber\\
&\left.\left.-11443047 r^7 \tilde{\alpha } r_h^4+77 r^6 r_h^5 \left(539773 \tilde{\alpha }-115200 \tilde{\sigma }\right)+33 r^5 r_h^6 \left(417457 \tilde{\alpha }-115200 \tilde{\sigma }\right)\right.\right.\nonumber\\
&\left.\left.+231 r^4 r_h^7 \left(68821 \tilde{\alpha }-16200 \tilde{\sigma }\right)-616 r^3 r_h^8 \left(117379 \tilde{\alpha }-22800 \tilde{\sigma }\right)-1848 r^2 r_h^9 \left(7159 \tilde{\alpha }-1620 \tilde{\sigma }\right)\right.\right.\nonumber\\
&\left.\left.-420 r r_h^{10} \left(34091 \tilde{\alpha }-7200 \tilde{\sigma }\right)+69300 r_h^{11} \left(803 \tilde{\alpha }-84 \tilde{\sigma }\right)+5028639 r^{11} \tilde{\alpha }\right)+114 r^{12} \left(64461 \tilde{\alpha }+19060 \tilde{\sigma }\right)\right)\,,\\[2ex]
B_5(r)=&~\frac{2 \tilde{\alpha }^3 r_h}{45045 r^{15} \left(r-r_h\right)} \left(r^{15} \left(4 \tilde{\alpha } \left(488014 \tilde{\gamma }+130309 \tilde{\kappa }\right)-119235 \tilde{\eta }\right)-r_h^3 \left(-2648646 r^{11} \tilde{\alpha } \tilde{\gamma } r_h-18018 r^{10} \tilde{\alpha } r_h^2 \right.\right.\nonumber\\
&\left.\left.\times\left(115 \tilde{\gamma }-12 \tilde{\kappa }\right)+6006 r^9 r_h^3 \left(\tilde{\alpha } \left(1381 \tilde{\gamma }+60 \tilde{\kappa }\right)-40 \left(12 \tilde{\gamma } \tilde{\sigma }+\tilde{\eta }\right)\right)+2574 r^8 r_h^4 \left(\tilde{\alpha } \left(2519 \tilde{\gamma }+1300 \tilde{\kappa }\right)-280 \tilde{\gamma } \tilde{\sigma }+60 \tilde{\eta }\right)\right.\right.\nonumber\\
&\left.\left.+9009 r^7 r_h^5 \left(4 \tilde{\alpha } \left(192 \tilde{\gamma }+85 \tilde{\kappa }\right)-80 \tilde{\gamma } \tilde{\sigma }+15 \tilde{\eta }\right)-8008 r^6 r_h^6 \left(4343 \tilde{\alpha } \tilde{\gamma }-355 \tilde{\alpha } \tilde{\kappa }-990 \tilde{\gamma } \tilde{\sigma }-195 \tilde{\eta }\right)\right.\right.\nonumber\\
&\left.\left.-18018 r^5 r_h^7 \left(\tilde{\alpha } \left(193 \tilde{\gamma }+332 \tilde{\kappa }\right)-80 \tilde{\gamma } \tilde{\sigma }+36 \tilde{\eta }\right)-1638 r^4 r_h^8 \left(\tilde{\alpha } \left(2757 \tilde{\gamma }+3620 \tilde{\kappa }\right)-880 \tilde{\gamma } \tilde{\sigma }+390 \tilde{\eta }\right)\right.\right.\nonumber\\
&\left.\left.+6006 r^3 r_h^9 \left(\tilde{\alpha } \left(9249 \tilde{\gamma }-980 \tilde{\kappa }\right)-15 \left(80 \tilde{\gamma } \tilde{\sigma }+31 \tilde{\eta }\right)\right)+1386 r^2 r_h^{10} \left(\tilde{\alpha } \left(179 \tilde{\gamma }+2020 \tilde{\kappa }\right)-520 \tilde{\gamma } \tilde{\sigma }+330 \tilde{\eta }\right)\right.\right.\nonumber\\
&\left.\left.+25740 r r_h^{11} \left(\tilde{\alpha } \left(13 \tilde{\gamma }+110 \tilde{\kappa }\right)-28 \tilde{\gamma } \tilde{\sigma }+18 \tilde{\eta }\right)+12012 r_h^{12} \left(\tilde{\alpha } \left(238 \tilde{\kappa }-2365 \tilde{\gamma }\right)+180 \tilde{\gamma } \tilde{\sigma }+119 \tilde{\eta }\right)+24024 r^{12} \tilde{\alpha } \tilde{\gamma }\right)\right)\,,\\[2ex]
B_6(r)=&~\frac{\tilde{\alpha}^3 r_h}{212270058000 r^{18} (r - r_h)} \left(6792641856000 \tilde{\gamma} \tilde{\eta} (2 r - 3 r_h) (r - r_h)^2 r_h^9 (r^2 + r r_h + r_h^2)^3 \right.\nonumber\\
&+ 1080 \tilde{\alpha}^2 \tilde{\sigma} \left(169018655174 r^{18} + 46232619433 r^{17} r_h + 21113485393 r^{16} r_h^2 - 37765164437 r^{15} r_h^3\right.\nonumber\\
&\left.- 71559054567 r^{14} r_h^4 - 76844109342 r^{13} r_h^5 - 54679534910 r^{12} r_h^6 - 242926912800 r^{11} r_h^7 - 342017706030 r^{10} r_h^8\right.\nonumber\\
&\left.+ 725632872965 r^9 r_h^9 + 622202730149 r^8 r_h^{10} + 955122256907 r^7 r_h^{11} - 2399335381443 r^6 r_h^{12} - 1098060215868 r^5 r_h^{13}\right.\nonumber\\
&\left.- 1471609807668 r^4 r_h^{14} + 3007888877994 r^3 r_h^{15} + 679039050690 r^2 r_h^{16} + 818189031660 r r_h^{17} - 1249641693300 r_h^{18}\right)\nonumber\\
&+ \tilde{\alpha}^3 \left(300971654678108 r^{18} + 185774588693694 r^{17} r_h + 149114714227479 r^{16} r_h^2 - 91851374550936 r^{15} r_h^3\right.\nonumber\\
&\left.- 561611162253141 r^{14} r_h^4 - 726292142175600 r^{13} r_h^5 + 548302343853264 r^{12} r_h^6 + 757385927956944 r^{11} r_h^7\right.\nonumber\\
&\left.+ 1167332248396314 r^{10} r_h^8 - 2964027148988906 r^9 r_h^9 - 1441662041696877 r^8 r_h^{10} - 2383010100934008 r^7 r_h^{11}\right.\nonumber\\
&\left.+ 8780232849214827 r^6 r_h^{12} + 1756504212588972 r^5 r_h^{13} + 2909072730786222 r^4 r_h^{14} - 11871788847574656 r^3 r_h^{15}\right.\nonumber\\
&\left.- 1647685324385100 r^2 r_h^{16} - 2013852740499600 r r_h^{17} + 7147089612663000 r_h^{18}\right)\nonumber\\
&- 2494800 \tilde{\alpha} \left(-21781760 \tilde{\gamma} \tilde{\kappa} (r - r_h)^3 r_h^7 (r^2 + r r_h + r_h^2)^4 + \tilde{\sigma}^2 \left(-11838611 r^{18} - 6223360 r^{11} r_h^7\right.\right.\nonumber\\
&\left.\left.- 7487480 r^{10} r_h^8 + 78278200 r^9 r_h^9 + 68136068 r^8 r_h^{10} + 79825200 r^7 r_h^{11} - 169829660 r^6 r_h^{12}\right.\right.\nonumber\\
&\left.\left.- 87703000 r^5 r_h^{13} - 106623660 r^4 r_h^{14} + 135863728 r^3 r_h^{15} + 29864835 r^2 r_h^{16} + 38238200 r r_h^{17} - 40500460 r_h^{18}\right)\right.\nonumber\\
&+ 4 \tilde{\gamma}^2 \left(635421 r^{18} - 3014440 r^{11} r_h^7 - 3233230 r^{10} r_h^8 + 7487480 r^9 r_h^9 - 5717712 r^8 r_h^{10}\right.\nonumber\\
&\left.\left.\left.- 5383560 r^7 r_h^{11} - 37777740 r^6 r_h^{12} + 19320840 r^5 r_h^{13} + 19253520 r^4 r_h^{14} + 51867816 r^3 r_h^{15}\right.\right.\right.\nonumber\\
&\left.\left.\left.- 10805795 r^2 r_h^{16} - 10850840 r r_h^{17} - 21781760 r_h^{18}\right)\right)\right)\,;
\end{align*}

\begin{align*}
\phi_1(r)=&~\frac{2 \tilde{\alpha } r_h \left(3 r r_h+2 r_h^2+6 r^2\right)}{3 r^3}\,,\\[2ex]
\phi_2(r)=&~0\,,\\[2ex]
\phi_3(r)=&~\frac{2 \tilde{\alpha }^2 r_h }{4725 r^9} \left(45990 r^7 \tilde{\alpha } r_h+55230 r^6 \tilde{\alpha } r_h^2+945 r^5 r_h^3 \left(11 \tilde{\alpha }+20 \tilde{\sigma }\right)-126 r^4 r_h^4 \left(119 \tilde{\alpha }-150 \tilde{\sigma }\right)\right.\nonumber\\
&\left.-420 r^3 r_h^5 \left(82 \tilde{\alpha }-45 \tilde{\sigma }\right)+450 r^2 r_h^6 \left(35 \tilde{\alpha }-6 \tilde{\sigma }\right)+945 r r_h^7 \left(29 \tilde{\alpha }-5 \tilde{\sigma }\right)+700 r_h^8 \left(53 \tilde{\alpha }-9 \tilde{\sigma }\right)-630 r^8 \tilde{\alpha }\right)\,,\\[2ex]
\phi_4(r)=&~\frac{\tilde{\alpha }^2 r_h^5}{3465 r^{12}} \left(44352 r^7 \tilde{\alpha } \left(\tilde{\gamma }+\tilde{\kappa }\right) + 73920 r^6 \tilde{\alpha } r_h \left(\tilde{\gamma }+\tilde{\kappa }\right) + 7920 r^5 r_h^2 \left(12 \tilde{\alpha } \tilde{\kappa }+\tilde{\eta }\right)\right.\nonumber\\
&+ 3465 r^4 r_h^3 \left(\tilde{\eta }-4 \tilde{\alpha } \left(\tilde{\gamma }-2 \tilde{\kappa }\right)\right) - 24640 r^3 \tilde{\alpha } r_h^4 \left(\tilde{\gamma }+\tilde{\kappa }\right) - 2772 r^2 r_h^5 \left(\tilde{\alpha } \left(24 \tilde{\kappa }-36 \tilde{\gamma }\right)+5 \tilde{\eta }\right)\nonumber\\
&\left.- 10080 r r_h^6 \left(\tilde{\alpha } \left(4 \tilde{\kappa }-8 \tilde{\gamma }\right)+\tilde{\eta }\right) - 2310 r_h^7 \left(\tilde{\alpha } \left(8 \tilde{\kappa }-28 \tilde{\gamma }\right)+3 \tilde{\eta }\right)\right)\,,\\[2ex]
\phi_5(r)=&~\frac{\tilde{\alpha }^3 r_h}{156080925 r^{15}} \left(36036 r^{13} \tilde{\alpha } r_h \left(705449 \tilde{\alpha }+261270 \tilde{\sigma }\right) + 6006 r^{12} \tilde{\alpha } r_h^2 \left(9815081 \tilde{\alpha }+2131500 \tilde{\sigma }\right)\right.\nonumber\\
&+ 18018 r^{11} \tilde{\alpha } r_h^3 \left(1914503 \tilde{\alpha }+252030 \tilde{\sigma }\right) + 36036 r^{10} \tilde{\alpha } r_h^4 \left(376751 \tilde{\alpha }+439458 \tilde{\sigma }\right)\nonumber\\
&- 12012 r^9 \tilde{\alpha } r_h^5 \left(2011573 \tilde{\alpha }-2284830 \tilde{\sigma }\right) + 15444 r^8 r_h^6 \left(681610 \tilde{\alpha } \tilde{\sigma }+906299 \tilde{\alpha }^2-184800 \left(\tilde{\gamma }^2-4 \tilde{\sigma }^2\right)\right)\nonumber\\
&+ 9009 r^7 r_h^7 \left(-950490 \tilde{\alpha } \tilde{\sigma }+5193317 \tilde{\alpha }^2-554400 \left(\tilde{\gamma }^2-3 \tilde{\sigma }^2\right)\right)\nonumber\\
&+ 14014 r^6 r_h^8 \left(-2495700 \tilde{\alpha } \tilde{\sigma }+7334099 \tilde{\alpha }^2-118800 \left(4 \tilde{\gamma }^2-11 \tilde{\sigma }^2\right)\right)\nonumber\\
&- 720720 r^5 r_h^9 \left(-37332 \tilde{\alpha } \tilde{\sigma }+57335 \tilde{\alpha }^2+693 \left(4 \tilde{\gamma }^2+5 \tilde{\sigma }^2\right)\right)\nonumber\\
&- 171990 r^4 r_h^{10} \left(-283080 \tilde{\alpha } \tilde{\sigma }+454159 \tilde{\alpha }^2-5280 \left(2 \tilde{\gamma }^2-9 \tilde{\sigma }^2\right)\right)\nonumber\\
&- 2522520 r^3 r_h^{11} \left(-30538 \tilde{\alpha } \tilde{\sigma }+54535 \tilde{\alpha }^2+495 \left(11 \tilde{\sigma }^2-4 \tilde{\gamma }^2\right)\right)\nonumber\\
&+ 426888 r^2 r_h^{12} \left(-2595 \tilde{\alpha } \tilde{\sigma }+112862 \tilde{\alpha }^2+1800 \left(4 \tilde{\gamma }^2-\tilde{\sigma }^2\right)\right)\nonumber\\
&+ 2162160 r r_h^{13} \left(-6778 \tilde{\alpha } \tilde{\sigma }+36286 \tilde{\alpha }^2+660 \tilde{\gamma }^2+495 \tilde{\sigma }^2\right)\nonumber\\
&\left.+ 27747720 r_h^{14} \left(-1056 \tilde{\alpha } \tilde{\sigma }+4049 \tilde{\alpha }^2+108 \tilde{\sigma }^2\right) - 18018 r^{14} \tilde{\alpha } \left(829843 \tilde{\alpha }+41340 \tilde{\sigma }\right)\right)\,,\\[2ex]
\phi_6(r)=&~\frac{4 \tilde{\alpha }^3 r_h}{11497961475 r^{18}} \left(-1021020 \tilde{\alpha } \left(\tilde{\alpha } \left(674536 \tilde{\gamma }-44204 \tilde{\kappa }\right)-59895 \tilde{\eta }\right) r^{17}\right.\nonumber\\
&+30630600 \tilde{\alpha } \left(4 \tilde{\alpha } \left(5323 \tilde{\gamma }+2356 \tilde{\kappa }\right)-989 \tilde{\eta }\right) r_h r^{16}\nonumber\\
&+340340 \tilde{\alpha } \left(4 \tilde{\alpha } \left(807394 \tilde{\gamma }+271669 \tilde{\kappa }\right)-178575 \tilde{\eta }\right) r_h^2 r^{15}\nonumber\\
&+15315300 \tilde{\alpha } \left(4 \tilde{\alpha } \left(5323 \tilde{\gamma }+2356 \tilde{\kappa }\right)-989 \tilde{\eta }\right) r_h^3 r^{14}\nonumber\\
&+2450448 \tilde{\alpha } \left(\tilde{\alpha } \left(103457 \tilde{\gamma }+44117 \tilde{\kappa }\right)-4945 \tilde{\eta }\right) r_h^4 r^{13}\nonumber\\
&+2042040 \tilde{\alpha } \left(4 \tilde{\alpha } \left(98687 \tilde{\gamma }+83852 \tilde{\kappa }\right)-4945 \tilde{\eta }\right) r_h^5 r^{12}\nonumber\\
&+2187900 \tilde{\alpha } \left(25073 \tilde{\eta }+\tilde{\alpha } \left(405488 \tilde{\gamma }+706364 \tilde{\kappa }\right)\right) r_h^6 r^{11}\nonumber\\
&+765765 \tilde{\alpha } \left(86063 \tilde{\eta }+4 \tilde{\alpha } \left(128591 \tilde{\gamma }+386780 \tilde{\kappa }\right)+600600 \left(\tilde{\gamma }+2 \tilde{\kappa }\right) \tilde{\sigma }\right) r_h^7 r^{10}\nonumber\\
&-340340 \tilde{\alpha } \left(-99625 \tilde{\eta }+4 \tilde{\alpha } \left(591952 \tilde{\gamma }+293077 \tilde{\kappa }\right)-2702700 \left(\tilde{\gamma }+2 \tilde{\kappa }\right) \tilde{\sigma }\right) r_h^8 r^9\nonumber\\
&+102102 \left(4 \left(7687741 \tilde{\gamma }-7086146 \tilde{\kappa }\right) \tilde{\alpha }^2-9 \left(547181 \tilde{\eta }+60060 \left(7 \tilde{\gamma }-50 \tilde{\kappa }\right) \tilde{\sigma }\right) \tilde{\alpha }+1441440 \tilde{\eta } \tilde{\sigma }\right) r_h^9 r^8\nonumber\\
&+556920 \left(4 \left(1842160 \tilde{\gamma }-379271 \tilde{\kappa }\right) \tilde{\alpha }^2+\left(-740477 \tilde{\eta }-180180 \left(7 \tilde{\gamma }-10 \tilde{\kappa }\right) \tilde{\sigma }\right) \tilde{\alpha }+180180 \tilde{\eta } \tilde{\sigma }\right) r_h^{10} r^7\nonumber\\
&+340340 \left(32 \left(516121 \tilde{\gamma }+219535 \tilde{\kappa }\right) \tilde{\alpha }^2-6 \left(131816 \tilde{\eta }+45045 \left(11 \tilde{\gamma }+10 \tilde{\kappa }\right) \tilde{\sigma }\right) \tilde{\alpha }+135135 \tilde{\eta } \tilde{\sigma }\right) r_h^{11} r^6\nonumber\\
&-10367280 \left(\left(608889 \tilde{\gamma }-649983 \tilde{\kappa }\right) \tilde{\alpha }^2+\left(-104906 \tilde{\eta }-20475 \left(13 \tilde{\gamma }-14 \tilde{\kappa }\right) \tilde{\sigma }\right) \tilde{\alpha }+34125 \tilde{\eta } \tilde{\sigma }\right) r_h^{12} r^5\nonumber\\
&-218790 \left(4 \left(7395637 \tilde{\gamma }-4195190 \tilde{\kappa }\right) \tilde{\alpha }^2+\left(-3863609 \tilde{\eta }-180180 \left(71 \tilde{\gamma }-50 \tilde{\kappa }\right) \tilde{\sigma }\right) \tilde{\alpha }+1441440 \tilde{\eta } \tilde{\sigma }\right) r_h^{13} r^4\nonumber\\
&-116804688 \left(\left(57226 \tilde{\gamma }+394 \tilde{\kappa }\right) \tilde{\alpha }^2+\left(630 \left(10 \tilde{\kappa }-37 \tilde{\gamma }\right) \tilde{\sigma }-4736 \tilde{\eta }\right) \tilde{\alpha }+2205 \tilde{\eta } \tilde{\sigma }\right) r_h^{14} r^3\nonumber\\
&+58963905 \left(4 \left(32272 \tilde{\gamma }-18665 \tilde{\kappa }\right) \tilde{\alpha }^2+\left(1170 \left(10 \tilde{\kappa }-11 \tilde{\gamma }\right) \tilde{\sigma }-16979 \tilde{\eta }\right) \tilde{\alpha }+1560 \tilde{\eta } \tilde{\sigma }\right) r_h^{15} r^2\nonumber\\
&+154594440 \left(92 \left(493 \tilde{\gamma }-218 \tilde{\kappa }\right) \tilde{\alpha }^2+\left(70 \left(50 \tilde{\kappa }-83 \tilde{\gamma }\right) \tilde{\sigma }-5451 \tilde{\eta }\right) \tilde{\alpha }+630 \tilde{\eta } \tilde{\sigma }\right) r_h^{16} r\nonumber\\
&\left.+340680340 \left(12 \left(1535 \tilde{\gamma }-398 \tilde{\kappa }\right) \tilde{\alpha }^2+\left(90 \left(10 \tilde{\kappa }-31 \tilde{\gamma }\right) \tilde{\sigma }-1933 \tilde{\eta }\right) \tilde{\alpha }+270 \tilde{\eta } \tilde{\sigma }\right) r_h^{17}\right)\,.
\end{align*}

\end{widetext}

\bibliography{biblio}
\end{document}